\documentclass{optica-article}

\journal{opticajournal} 

\articletype{Research Article}

\usepackage{lineno}
\usepackage{graphicx}
\usepackage{amsmath}
\usepackage{bm}
\usepackage[normalem]{ulem}

\begin{document}

\title{Direct 3D imaging through spatial coherence of light}

\author{Gianlorenzo Massaro,\authormark{1,2,*} Barbara Barile,\authormark{3} Giuliano Scarcelli,\authormark{4} Francesco V. Pepe,\authormark{1,2} Grazia Paola Nicchia,\authormark{3,5,6,$\dagger$} and Milena D'Angelo\authormark{1,2,$\dagger$}}

\address{
\authormark{1}Dipartimento Interateneo di Fisica, Università degli Studi di Bari Aldo Moro, Bari, 70125,Italy\\
\authormark{2}Sezione di Bari, Istituto Nazionale di Fisica Nucleare, Bari, 70125,Italy\\
\authormark{3}Dipartimento di Bioscienze, Biotecnologie e Ambiente,Università degli Studi di Bari Aldo Moro, Bari, 70125,Italy\\
\authormark{4}Fischell Department of Bioengineering, University of Maryland, College Park MD 20742 USA\\
\authormark{5}Institute for Organic Synthesis and Photoreactivity, National Research Council of Italy, Bologna, 40129, Italy\\
\authormark{6}Dominik P.Purpura Department of Neuroscience, Albert Einstein College of Medicine, New York, 10461, NY, USA\\
\authormark{$\dagger$}The authors contributed equally to this work.
}

\email{\authormark{*}gianlorenzo.massaro@uniba.it} 


\begin{abstract*} 
Wide-field imaging is widely adopted due to its fast acquisition, cost-effectiveness and ease of use. Its extension to direct volumetric applications, however, is burdened by the trade-off between resolution and depth of field (DOF), dictated by the numerical aperture of the system. 
We demonstrate that such trade-off is not intrinsic to wide-field imaging, but stems from the spatial incoherence of light: images obtained through spatially coherent illumination are shown to have resolution and DOF independent of the numerical aperture.
This fundamental discovery enabled us to demonstrate an optimal combination of coherent resolution-DOF enhancement and incoherent tomographic sectioning for scanning-free, wide-field 3D microscopy on a multicolor histological section.
\end{abstract*}

\section{Introduction}
Wide-field imaging is amongst the most common imaging modalities for the observation and characterization of absorbing specimens, as done, for instance, in bright-field microscopy \cite{microscopy_ch1}. Some of the reasons behind its widespread use across many diverse applications are its ease of use, cost-effectiveness, fast acquisition, and its {\it direct} imaging capability (namely, the availability of the output image in real time, with no need for inverse computation techniques on the collected intensity).
Although conventional devices work extremely well with 2D samples, having negligible thickness along the optical axis ($z$), their use with 3D samples is significantly complicated by the well-known dependence of both resolution and depth of field (DOF) on the numerical aperture (NA) of the imaging device: this dependence results in a strong trade-off between image resolution and DOF, and imposes the need to \textit{z-scan} the whole sample in order to collect the complete volumetric profile. 
The operation of z-scanning requires that either the imaging device or the sample itself are mechanically shifted along the optical axis, so as to change the plane at focus and perform multiple acquisitions of different transverse planes \cite{3dMicMethods, 3Dmicroscopy}.
The intrinsically long acquisition required by moving components implies limited {\it in vivo} applicability and comes with further disadvantages, such as the need for precise stabilization, requiring large and heavy devices, costly mechanical parts with the required precision, as well as high maintenance costs, which preclude the use of scanning microscopes in low-budget applications.
The limitations of axial scanning become particularly relevant in large-NA devices, where the higher resolution comes at the expense of a narrower DOF. This has detrimental effects on the number of axial measurements necessary to characterize the entire sample, so that a common option to keep the measurement time low is to under-sample along the optical axis, with a consequent loss of information.
In 3D imaging, resolution, axial sampling and acquisition speed are thus in direct conflict.

Several apporaches have been proposed in the literature to address this problem; optical coherence tomography (OCT) is one of the most noticeable examples \cite{OCT}. However, in all cases, the limitations imposed by the NA of the imaging system persist; in OCT, for example, small NA are required, at the expenses of resolution and signal-to-noise ratio (SNR), for addressing the loss of intensity implied by large-NA optics \cite{ELADAWI2020191}.

A recent and rapidly developing approach to scanning-free wide-field 3D microscopy is light-field (LF) imaging, where direct images of thick samples containing heavily defocused planes are acquired and then {\it refocused}, in post-processing  \cite{adelson1992single,Ng2005LightFP, 10.1145/1179352.1141976,levoy2006light,broxton2013wave}. Directional information about light from the sample is in fact acquired by a microlens array and employed, in post-processing, to perform software z-scans with similar features to the typical mechanical scans.
LF devices thus enable scanning-free single-shot acquisition of a 3D sample, but its fast acquisition comes at the expenses of a dramatic loss of resolution, well beyond the diffraction limit \cite{PI6}. 
In fact, due to its geometric-optics-based working principle, the maximum achievable DOF is defined by the circle of confusion (CoC), namely, by the projection of the lens aperture over the acquired defocused planes \cite{Stokseth:69}. The resolution of the refocused images is thus not determined by the Airy disk, as is typically the case in microscopy, but is rather dominated by the geometrical effects of defocusing, as typically occurring in photography.
In addition, the lenslets introduce an even stronger trade-off between resolution and DOF, consisting in the loss of resolution at focus with the improvement of the volumetric performance \cite{inproceedings}.


In this work, we demonstrate that the resolution versus DOF trade-off of defocused images is governed by the {\it spatial} coherence properties of light, and is naturally relieved when the sample is illuminated with spatially coherent light (\textit{i.e.}, the coherence area on the sample is either comparable or larger than the sample details, as explained in the Methods and Results). {\it Coherent} imaging is thus found to entail a much slower image degradation with defocusing, a result that leads to discover the direct 3D imaging potential of spatially coherent light: by combining the extremely large DOF of coherent imaging with the strong localization capability of incoherent imaging, we design a direct, scanning-free, wide-field 3D microscope and demonstrate its working principle by means of both test and histological samples. In particular, we characterize the properties of direct coherent wide-field imaging and show 3D reconstruction compatible with absorbing non-fluorescent dyes routinely used for histochemistry~\cite{staining}.

An important aspect to remark is that the required coherence exclusively relates with the transverse coherence of the field illuminating the sample, disregarding both the temporal and the spatial coherence of the source \cite{saleh.ch10}: although our findings apply to both temporally and spatially coherent sources such as lasers, as well as to collimated beams, none of these properties are necessary to our scopes, and our results are thus not confined to these scenarios. On the contrary, NA-independent resolution and DOF are shown to be obtained with virtually any source of spatially and temporally incoherent light, such as a LED, since the required spatial coherence can always be acquired through propagation (Van Cittert-Zernike theorem \cite{mansuripur_2009}).

Our 3D microscope is, in fact, 
based on a conventional bright-field imaging device, integrated with an array of LEDs for implementing a dedicated coherent illumination strategy. Conversely, typical bright-field illumination is obtained with extended sources, shining spatially incoherent light on the sample \cite{hecht2012optics, microscopy_ch1}. 
Spatial coherence, on the other hand, is used in a plethora of non real-time imaging modalities relying on post-processing of the acquired data aimed at recovering rich phase information about the sample; techniques such as holography \cite{holo1949, alma9926533469005776} and ptychography \cite{Rodenburg2019}, for example, can achieve super-resolution, wavefront reconstruction, and correction of optical aberrations. Most notably, techniques based on computational illumination from LED arrays \cite{FourPtych_2021, Zheng:11} have demonstrated high-resolution 3D amplitude and phase reconstruction \cite{Tian:15} by exploiting sequential multi-angle plane-wave illumination and recursive phase-retrieval algorithms.
However, all such coherent imaging techniques are indirect, due to the time-consuming algorithms they require for data analysis, and are thus not suitable for real-time imaging \cite{8735704,holo_recons}.

In this work, we show that the spatial coherence of light can be exploited in direct wide-field imaging to obtain a breakthrough improvement of the image resolution over large DOF. 
This result is supported by the discovery of the completely different physical mechanisms regulating resolution loss in defocused images obtained through spatially coherent and spatially incoherent illumination.
In fact, while the peculiarities of focused images, whether coherent or incoherent, are well known \cite{Aert:06, goodman2005introduction}, the properties of coherent defocused images have been so far mostly unexplored, with the only exception of the very special case of collimated light illumination \cite{Tian:14,Tian:15, goodman2005introduction}.
In this work, the introduction of a dedicated formalism and an unbiased image quantifier enables to study the properties of coherent images and to compare them with the ones of conventional incoherent imaging. 
One of the main results we shall present is that neither the NA nor the design of the imaging system affect the quality of defocused coherent images; in fact, the NA-dependent trade-off between resolution and DOF defined, in incoherent defocused images, by the CoC, naturally disappears when illuminating the sample with spatially coherent light. We shall profit from this effect to perform the typical tomographic reconstruction of LF imaging and retain its multicolor capability, but with enhanced resolution both at focus (where we recover Rayleigh-limited resolution) and in refocused planes. No phase retrieval and time-consuming post processing of the acquired images are required in our approach, paving the way toward 3D real-time imaging.


\section{Methods}

As mentioned in the Introduction, we refer to \textit{coherent} imaging whenever the coherence area of the illumination \cite{scully_zubairy_1997}, {\it on the sample}, is larger than the spatial features of the sample one wishes to resolve \cite{saleh.ch10}.
According to the size of the details composing a given object, an imaging system might thus behave \textit{coherently} for object details smaller than the coherence area, and \textit{incoherently} for larger details. 
In this respect, we should highlight that the size of the coherence area with respect to the whole field of view (FOV) of the image does not play any role.
The transition from one regime to the other will be discussed in details later in the paper. For the sake of simplicity, we shall now disregard the effects of partial coherence, and only consider coherent systems as having a coherence area larger than any object detail, and incoherent systems as having a negligible (point-like) coherence area on the sample.

\subsection{Resolution and DOF in coherent imaging}

\begin{figure}
    \centering
    \includegraphics[width=\textwidth]{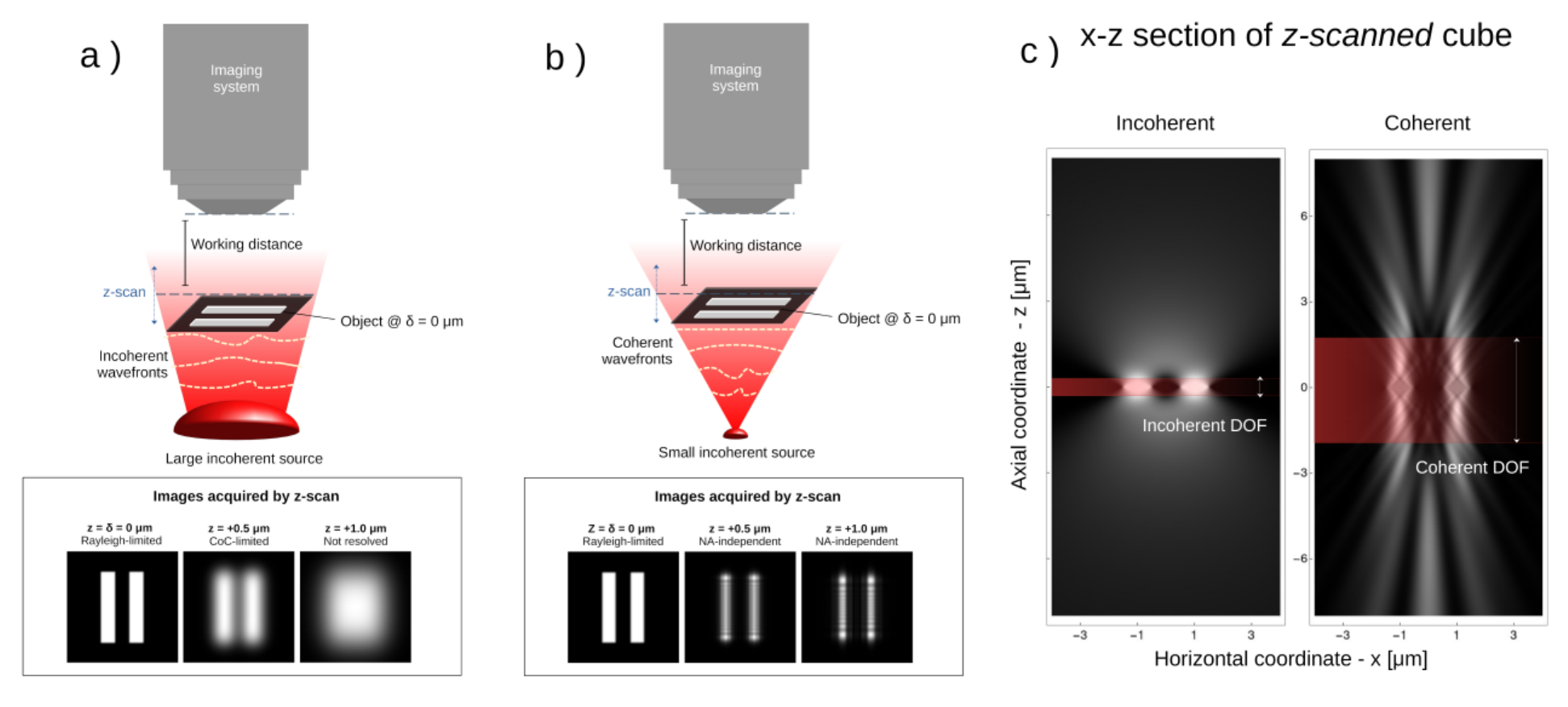}
    \caption{\label{fig:fig1}
    \textbf{Comparison of resolution versus DOF in incoherent and coherent imaging.}
    \textit{Panel a).} Incoherent imaging setup (top) and corresponding focused and defocused images (bottom). In the setup, a transmissive sample (a double-slit mask) is illuminated by the incoherent wavefronts coming from an extended source and is imaged by a conventional imaging system. The focused image (bottom left) is obtained by placing the sample at the working distance of the imaging system; the defocused images (bottom center and right) are obtained by $z$-scanning of the sample.
    \textit{Panel b).} Coherent imaging setup (top) and corresponding focused and defocused images (bottom). In the setup, the same transmissive sample as in panel a) is illuminated by spatially coherent light from a small incoherent source, and is imaged by the same conventional imaging system. The focused (bottom left) and defocused (bottom center and right) coherent images are obtained by placing the sample at exactly the same positions as in panel a).
    \textit{Panel c).} Axial section of the 3D cube obtained by z-scanning the sample in the setups of panels (a) and (b), namely, in the case of incoherent (left panel) and coherent (right panel) illumination, demonstrating the extremely larger DOF of coherent imaging.
    }
\end{figure}

The upper part of panel b) suggests one of many possible ways for obtaining coherent illumination from an incoherent source: since the coherence area on the sample scales proportionally with the ratio between the source diameter and the source distance, the desired coherence can easily be obtained by reducing the source size \cite{mansuripur_2009}. An obvious alternative would be to employ laser light illumination, but the presented results are not limited to this scenario.
In Figs.~\ref{fig:fig1} a) and b), we report two typical examples of incoherent and coherent imaging, respectively, as obtained by changing the illumination in the same exact imaging system.
The opposite situation of incoherent illumination is typically achieved by placing extended natural sources at a small distance from the sample, as reported in the upper part of panel a).
In the lower part of panels a) and b), we report the corresponding incoherent and coherent images, both focused (left panels) and defocused (central and right panels), of a two-dimensional sample (a double-slit mask). Defocused images have very distinctive features depending on the spatial incoherence or coherence of the light on the sample: while incoherent images tend to quickly blur upon defocusing, coherent images do not blur. This is even more apparent in panel c), reporting a section, in the $(x, z)$ plane, of the 3D cubes obtained by mechanically z-scanning the two-slit mask, in both cases of incoherent (left panel) and coherent (right panel) imaging. Whereas, in incoherent imaging, z-scanning quickly gives rise to flat intensity distributions as the object is moved out of focus, in coherent imaging, the transmissive details contain rich spatial modulations and stay well separated from each other over a much longer axial range compared to the corresponding incoherent image. Transmissive details thus appear ``resolved'' at a much larger distance from the plane at focus, before being completely altered by diffraction
. A much longer DOF (or, equivalently, higher resolution of defocused images) is thus expected in coherent imaging, with image degradation not due to blurring. Upon quantitatively describing these effects, we shall find that resolution and DOF of defocused coherent images are actually completely independent of the NA of the imaging system.

The differences between coherent and incoherent systems can be traced back to the different underlying image formation processes, as formally expressed by the intensity distributions describing the images\cite{goodman2005introduction}:
\begin{align}
I_\text{inc}(\bm x)= 
\left\lvert \mathcal A(\bm x)\right\rvert^2 *
\left\lvert \mathcal P\left(\bm x \right)\right\rvert^2
\ \ \ \ \ \ \ \ 
I_\text{coh}(\bm x)= \left\lvert
\mathcal A(\bm x)*
\mathcal P\left(\bm x\right)\right\rvert^2
,
\label{eq:coh_vs_inc}
\end{align}
where $\mathcal A(\bm x)$ is the complex transmission function of the object, $\mathcal P(\bm x)$ is the Green's function describing the field propagation through the optical system, and $f*g$ denotes the convolution between two complex-valued functions, $f$ and $g$.
Unlike the incoherent image formation process, which is linear in the \textit{optical intensity}, coherent imaging is non-linear with respect to the object $\mathcal A (\bm x)$.
Therefore, although the same quantities are involved in both intensity distributions of Eq.~\eqref{eq:coh_vs_inc}, those contributing to the incoherent image formation are real and positive, whereas coherent imaging is sensitive to both the amplitude and phase of complex functions describing both the field distribution within the sample and its propagation through the imaging system \cite{alma9926533469005776, Rodenburg2019, FourPtych_2021, Tian:15}. In fact, upon neglecting optical aberrations, the \textit{coherent} (\textit{i.e.} complex) PSF $\mathcal P(\bm x)$ of Eq.~\eqref{eq:coh_vs_inc} can be decomposed into two contributions: 
\begin{equation}
   \mathcal P (\bm x) = \mathcal D_{z-\delta} (\bm x) * \mathcal P_0(\bm x),
  \label{eq:PSF}
\end{equation}
where $\mathcal P_0$ is the complex PSF describing the focused coherent image and determining the well-known Airy disk \cite{microscopy_chap5}, and $\mathcal D_{z-\delta}$ represents the field propagation over a distance $z-\delta$, with $\delta$ and $z$ the axial coordinates of the object point and of the plane at focus, respectively.
Depending on the placement of the sample and the numerical aperture of the device, the quality of the output image can thus be dominated either by the effects of out-of-focus propagation or by the Airy disk, with the two effects blending into each other only when the object is placed close to (but not perfectly on) focus. 
The corresponding transition between the focused and defocused image is well known in incoherent imaging: at focus, both the resolution ($\lambda /$NA) and the DOF ($\lambda /$NA$^2$) are determined by wave optics (Airy disk), with $\lambda$ the illumination wavelenght. However, as the object is moved outside of the natural DOF of the focused device, the PSF $\mathcal P$ is dominated by geometrical optics effects and reduces to the circle of confusion (namely, the projection of the lens aperture onto the defocused image plane) \cite{Stokseth:69}, which induces a typically circular blurring with a radius proportional to both the defocusing $|z-\delta|$ and the effective lens radius. 

The different physics regulating coherent and incoherent imaging helps developing an intuition about the different behaviour observed in Fig. \ref{fig:fig1}, but does not suffice to quantitatively compare the resolution and DOF of coherent and incoherent imaging. In fact, image quality estimators typically used for characterizing imaging performance, from two-point resolution criteria, such as Rayleigh's and Abbe's \cite{microscopy_chap6, Aert:06}, to more advanced ones, such as modulation transfer functions \cite{MTF}, all rely on the linearity of the (incoherent) image formation and the positiveness of the PSF, and thus fail in assessing the performance of coherent imaging.
For instance, the definition of a Rayleigh criterion prescribes that, because of the broadening effect of the \textit{incoherent} PSF, the image of two ''points'' is the superposition of two disks (Airy disks, at focus, CoC, out of focus). The resolution is then easily defined by arbitrarily setting an acceptable threshold to when the two disks are perceived as separated. But these methods cannot be applied as effectively to a non-linear process such as the coherent image formation, since coherence induces the appearance of spurious spatial frequency components.
Therefore, neither an approach based on modulation transfer functions, that require the harmonic content to be unaltered, nor the two-point visibility, which requires a relative minimum separation between the images of two points, can be used.

To quantify the performance of coherent and incoherent imaging systems, we thus introduce a general-purpose quality estimator: the functional $F_A$, which we shall refer to as \textit{image fidelity}, defined as a positive quantity $F_A[I(\bm x)]$ that compares the intensity distribution $I(\bm x)$ of the image produced by an imaging system directly with the original intensity profile of the object $A=\lvert\mathcal A\rvert^2$, namely,
\begin{equation}
	F_A[I]=\int\sqrt{A\left(\frac{\bm x}{M}\right) I(\bm x)}\,d\bm x,
	\label{eq:fidelity}
\end{equation}
where $M$ is the magnification of the imaging system in its plane at focus. Both $A$ and $I$ are normalized quantities for the definition of the fidelity to be consistent and to saturate to unity in the ideal case of perfect imaging ($I=A$).
Being completely independent of any detail of the image formation process, the fidelity enables performing image quality evaluation through any imaging device, as long as the shape of the known reference object is known: resolution and DOF shall thus be defined as the minimum object size and the maximum axial range producing a ``faithful'' image, as identified by a threshold set to the fidelity. Both these definitions apply equally well to focused and defocused images, thus enabling to study how resolution changes with defocusing.
Since incoherent imaging is only sensitive to the \textit{intensity} transmitted by the sample, our study will now be restricted to non-diffusive objects and will disregard phase information, namely, we shall consider field transmission profiles with $\mathrm{arg}\left(\mathcal A\right)=0$ uniformly in the sample, so that $\mathcal A = \left\lvert \mathcal A\right \rvert \geq 0$.

\begin{figure}
    \centering
        \includegraphics[
        trim={0 0 0 1.5cm},clip,
        width=1.\textwidth]{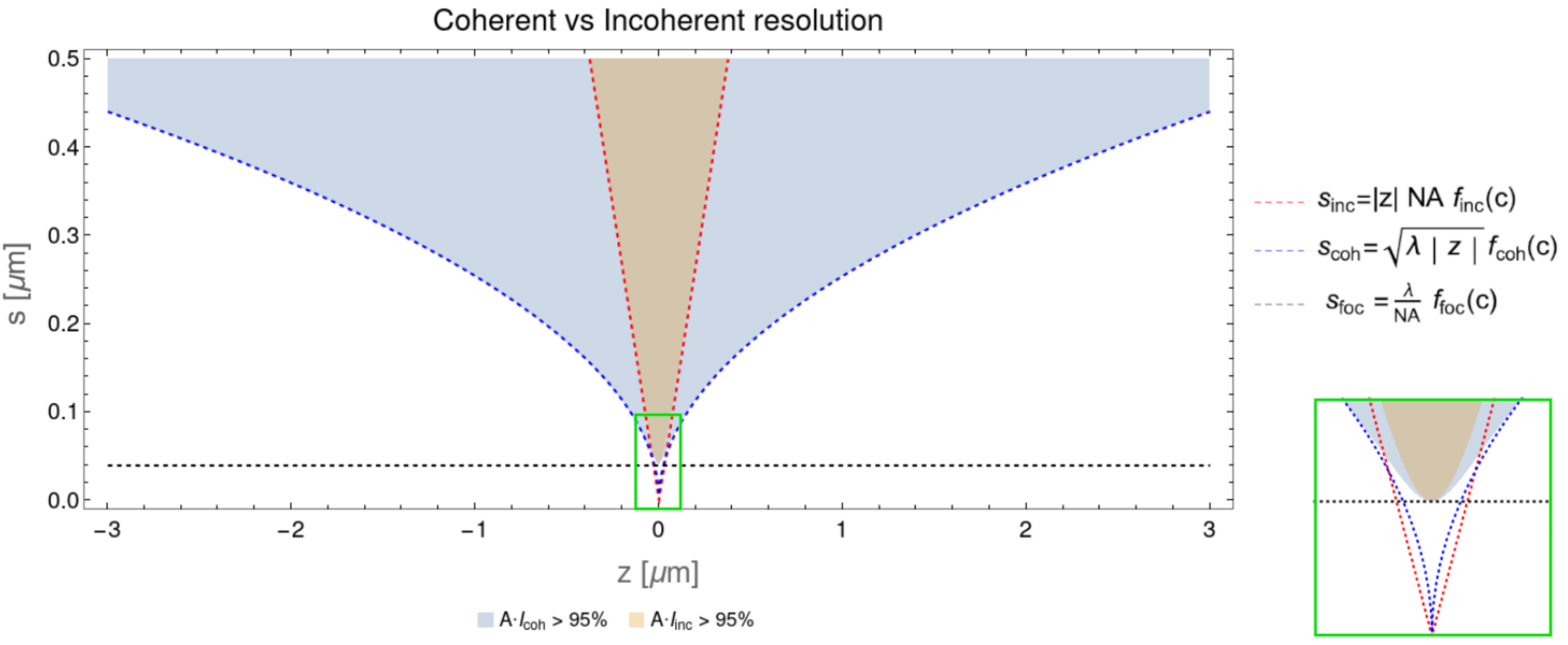}
\caption{\label{fig:coh_vs_inc}
    \textbf{Image fidelity in coherent and incoherent imaging.}
    The colored areas (blue for coherent illumination, orange for incoherent illumination) highlight the regions in which an $s$-sized object placed at an axial coordinate $z$ can be imaged with a fidelity larger than $95\%$, with $z=0$ identifying the position of the object at focus.
    The black dashed line is the fidelity equivalent of the Rayleigh criterion, intended as the minimum object size that can be imaged faithfully.
    The orange dashed curve is the NA-dependent geometrical circle of confusion at $95\%$ fidelity, obtained by evaluating the fidelity on the geometrical optics approximation. The blue dashed line represents the curve of $95\%$ fidelity obtained in the wave optics regime (namely, by considering the free space coherent propagation of the field) in the case of infinite NA of the imaging system.
    The imaging system is a $20 \times$ microscope with NA$=0.5$, illuminated with monochromatic light with wavelength $500\,$nm. The object is a mask with a single transmissive detail with Gaussian shape of width $s$.
    }
\end{figure}

\subsection{Resolution limits}

The plot reported in Fig.~\ref{fig:coh_vs_inc} employs the \textit{fidelity} to offer a quantitative interpretation of the coherent and incoherent $z$-scans reported in Fig.~\ref{fig:fig1}c). 
The colored areas in Fig.~\ref{fig:coh_vs_inc} highlight how far from the plane at focus (abscissas) an $s$-sized object (ordinates) can be placed to produce an image with fidelity higher than $95\%$. The orange area refers to spatially incoherent illumination, whereas the blue area refers to the coherent case.
The physical regimes leading to the dashed curves that delimit the high-fidelity regions associated with coherent and incoherent imaging offer a clear perspective on the physical mechanisms regulating the two image formation processes and enable quantifying the resolution versus DOF trade-off in the two cases. 
Such boundaries can be interpreted as \textit{resolution limit curves}, giving the functional dependence of the resolution on the displacement of the sample from focus, at the threshold of the image fidelity above which an image is considered resolved. 
These curves are obtained from the analytical expression of the image fidelity, written in terms of the parameters on which the image depends, which in our case are: the dimension $s$ of the features of the sample, the axial coordinate $\delta$ where the sample is located, and the axial location $z$ of the plane focused by the imaging system. The image fidelity associated with $I(\bm x)=I(\bm x; \delta-z, s)$ is thus a two-variable function: $F_A[I](\delta-z,s)$. Since the quality of the image, upon mechanical z-scanning, only depends on the relative distance between the object and the focused plane, we shall set for simplicity $\delta=0$ and interpret $z$ as the relative \textit{defocusing} distance.

By studying the analytical expression of $F_A[I](z, s)$, exact expressions of relevant image quantifiers can be extracted. For instance, $F_A[I](0, s)$ gives the image fidelity in the plane with Rayleigh-limited resolution, as a function of the object size. By inversion, one obtains, both for coherent and incoherent imaging,
\begin{equation}
     s_\text{foc} = \frac{\lambda}{\text{NA}}\,f_\text{foc}(c),\label{eq:atFocus}
\end{equation}
where $f_\text{foc}(c)$ is a coefficient depending of the threshold image fidelity, amounting to $0.157$ for $c=0.95$ (Fig.~\ref{fig:coh_vs_inc}).
Apart from the multiplying constant, which only depends on the arbitrary choice of a threshold on the fidelity, the equation corresponds to the well-known diffraction-limited resolution of focused imaging systems (dashed black line in Fig.~\ref{fig:coh_vs_inc}), as determined by the Airy disk. 
Therefore, the analysis in terms of fidelity recovers the well-known fact that the optical performance of focused coherent and incoherent systems is analogous.

The differences between the two illumination strategies emerge when investigating defocused images in two different physical regimes.
The geometrical optics regime 
is explored by considering the fidelity in the limit $\lambda\rightarrow 0$, namely,
\begin{equation}
	F_{\text{geom}}[I](z, s)=\lim_{\lambda\rightarrow 0}F_A[I].
\end{equation}
If this physical regime is investigated in the incoherent imaging case, the implicit curves $F_\text{geom}[I_\text{inc}]=c$, in the $(z,s)$ plane, have an explicit expression, which, unsurprisingly, prescribes the well-known circle of confusion of geometrical optics:
\begin{equation}
 s_\text{geom}(z) = \text{NA}\,\lvert z\rvert\,f_{\text{geom}}(c),\label{eq:CoC}
\end{equation}
with $f_\text{geom}(0.95)=1.97$. As shown in Fig.~\ref{fig:coh_vs_inc}, the CoC-defined trend perfectly traces the boundary of the fidelity area. Hence, the fidelity analysis confirms that wave optics has negligible effects on the optical performance of an incoherent system when the sample is moved away from perfect focus.
By exploring the same physical limit in the case of coherent imaging, the obtained analytical expression does not describe any physically relevant situation and does not have a counterpart in the shape of the fidelity region. 

Instead, interesting results are obtained, in the coherent case, by investigating the opposite regime, namely by neglecting geometrical effects. This is done by considering the radius of the limiting aperture $l \rightarrow \infty$, so as to completely ignore the influence of the imaging device. This condition is equivalent to considering an imaging system where the image formation process is solely governed by diffraction, from the object plane up to the plane at focus; in fact, in Eq. \eqref{eq:PSF}, $\mathcal P(\bm x)\rightarrow\mathcal D_z(\bm x)$, indicating that no CoC exists in this case. 
Upon setting a threshold $c$ to the fidelity of a coherent system with infinite NA 
\begin{equation}
F_\text{diff}[I_\text{coh}]=\lim_{l\rightarrow\infty}F_A[I_\text{coh}],\label{eq:cohFid}
\end{equation}
we obtain the resolution limit curves
\begin{equation}
s_\text{diff}(z)=\sqrt{\lambda \lvert z\rvert}\,f_{\text{diff}}(c),
\label{eq:diffraction}
\end{equation}
with $f_{\text{diff}}(0.95)=0.396$. Rather surprisingly, this square-root scaling of the resolution with defocusing perfectly reproduces the boundary of coherent imaging out of the plane at focus, as reported by the blue dashed line in Fig.~\ref{fig:coh_vs_inc}. As in the previous case, exploring the same physical limit in the case of incoherent illumination yields no interesting conclusion.
The optical performance of coherent and incoherent imaging are thus defined by two entirely different processes: the geometrical CoC (hence, the system NA) is basically the only factor limiting the resolution of defocused incoherent imaging; on the contrary, the aperture size and optical design of the imaging system play no role in coherent imaging, where the sole responsible for image degradation is diffraction and free-space space propagation from the object to the observation plane.
The different physical phenomena governing image degradation (geometric optics, as opposed to diffraction and wave propagation) have surprising effects on the image quality. Resolution and DOF of coherent imaging are found to be independent of the NA of the imaging system, and their trade-off is extremely relieved with respect to incoherent imaging, as defined by the square-root law (dashed blue line) compared to the linear dependence (dashed red line).

\subsection{Coherent 3D imaging with incoherent sectioning capability}
The newly discovered properties of direct coherent images can be integrated with the strong axial localization capability of incoherent imaging to achieve scanning-free 3D wide-field imaging of absorbing samples with enhanced volumetric resolution.
In fact, spatially coherent illumination will enable a (NA-independent) square-root scaling of transverse resolution, thus offering high lateral resolution over a long DOF; at the same time, the axial sectioning typical of spatially incoherent illumination entails, in the wide DOF accessed because through coherence, a precise sectioning capability, as enabled by large-NA tomographic systems \cite{dnumerical}. 
We should clarify that, in this context, the concepts of DOF and axial resolution are rather distinct: while the DOF represents the axial length of the volume where object details of a given size can be faithfully imaged, the axial resolution represents the axial sectioning capabilities, that is, how finely transverse planes within the DOF can be isolated along the axis.

The underlying principle for achieving high-resolution 3D imaging within a direct wide-field coherent system is similar to LF imaging: in both cases, information about the propagation direction of light enables scanning-free volumetric reconstruction. However, while LF imaging acquires the required directional information by means of the microlens array, our proposal prescribes to do it with spatially coherent illumination of the sample from different locations. Our approach will be shown to come with two major advantages: a much larger DOF, as defined by the square-root (as opposed to linear) scaling of the resolution with defocusing, and Rayleigh-limited images at focus. 

In the proposed scheme, 3D information about the sample is acquired by accessing the 4D function $I(\bm x_0,\bm x)$, where $\bm x_0$ is the transverse coordinate of a point-like emitter enabling spatially coherent illumination of the sample, and $\bm x$ is the transverse coordinate of the collected image. Sampling of the complete 4D function is performed by sequentially sweeping an illumination plane made of point-like emitters centered in $\bm x_0$, and collecting, for each coordinate $\bm x_0$, the resulting intensity 
\begin{equation}
    I(\bm x_0,\bm x) = \left\lvert
\left[\mathcal L_{\bm x_0}(\bm x)\mathcal A(\bm x)\right]*
\mathcal P\left(\bm x\right)\right\rvert^2,
\label{eq:plenopticFunc}
\end{equation}
where $\mathcal A$ and $\mathcal P$ are the same object transmittance and coherent PSF as in Eq. \eqref{eq:coh_vs_inc}, and $\mathcal L_{\bm x_0}$ is the Green's function propagating the field from the point-like source centered in $\bm x_0$ to the sample plane. 
As we shall discuss in the ``Results'' section, the wide freedom in the choice of $\mathcal L_{\bm x_0}$ (hence, of the illumination scheme) enables to greatly customize the optical performances of the proposed 3D imaging system.
Specifically, in order to encode 3D information into $I(\bm x_0,\bm x)$, illuminating the sample from many different \textit{angles} is not necessary. In previous works (see, \textit{e.g.}, Ref.~\cite{Tian:15}), in fact, $\mathcal L_{\bm x_0}$ has always been arranged in such a way to have an illumination distance $L$ between a source at coordinate $\bm x_0$ and the sample, such that the latter can be considered to be illuminated by tilted plane waves, corresponding to the choice
\begin{equation}
    \mathcal L_{\bm x_0}(\bm x) = \exp\left[
            i\,\frac{2\pi}{\lambda}\frac{\bm x_0}{L}\cdot \bm x
    \right],
\end{equation}
as conventionally done in tomographic systems.
However, as we shall discuss in the ``Results'' section, our complete formal analysis, and the consequent understanding of coherent imaging, enable to demonstrate that neither the angular illumination nor the requirement of collimated light are in any way necessary to encode 3D information into $I(\bm x_0,\bm x)$. Most importantly, understanding the underlying physics of coherent and incoherent imaging is the key for achieving scanning-free direct 3D imaging, with no need for time-consuming phase retrieval algorithms.

The intensity distribution described by Eq. \eqref{eq:plenopticFunc} is easily recognized as a coherent image, as in Eq. \eqref{eq:coh_vs_inc}, with the only difference that the object is now replaced by the expression $\mathcal L_{\bm x_0}\,\mathcal A$, emphasizing the role of the illumination scheme and the wide freedom in its design. The acquired 4D intensity can thus be expected to have mostly the features we have attributed to coherent images, such as the decoupling of the lateral resolution and DOF. However, the large DOF entails the lack of axial localization: thick 3D samples are imaged with high transverse resolution, but lack any axial localization.
To address the issue, we shall integrate the proposed technique with the properties of incoherent imaging, in which the tomographic properties are defined by the angular acceptance of the lens. The analogy with incoherent systems can easily be understood by considering the image resulting from the sum of the coherent images obtained from different illumination coordinates, namely,
\begin{equation}
    R_0(\bm x)=\int I(\bm x_0,\bm x)\,d\bm x_0 = I_\text{inc}(\bm x).
    \label{eq:Radon_focus}
\end{equation}
This equality can be analytically verified by plugging the expression of the plane-wave illumination into Eq. \eqref{eq:plenopticFunc} and integrating the result; however, any other illumination schemes presented in this work yields the same result. In fact, from a more intuitive standpoint, we can recognize that integrating over the entire illumination plane is equivalent to shining uniform \textit{incoherent} light onto the sample, which is exactly the typically sought-after experimental condition of uniform illumination in conventional systems (\textit{e.g.} Kohler illumination). It is thus not surprising that the integration must yield exactly the same results as conventional (incoherent) imaging: Rayleigh-limited resolution at focus, CoC blurring out of focus, and dependence on NA, with the only difference that the uniform illumination is achieved in post-processing. 
However, the mere integration reported in Eq. \eqref{eq:Radon_focus} is a rather poor way of employing the much larger amount of information contained within $I(\bm x_0,\bm x)$: due to the shallow DOF of incoherent imaging, the features of the sample are rapidly lost as it is moved away from perfect focus, more so for large NA. On the contrary, a Radon transformation of $I(\bm x_0,\bm x)$ (here expressed as a line integral):
\begin{equation}
    R_z(\bm x^\prime)=\int_{\gamma(\bm x^\prime)} I(\bm x_0,\bm x)\,dl,
    \label{eq:Radon}
\end{equation}
enables localizing the object within the much larger DOF characterizing coherent imaging. In Eq. \eqref{eq:Radon}, $\gamma(\bm x^\prime)$ are two lines of equations $\sin{\theta(z)} \bm x_0+\cos{\theta(x)} \bm x=\bm x^\prime$ defined in the spaces $(x_0,x)$ and $(y_0,y)$. In fact, the Radon transform $R_z(\bm x^\prime)$ isolates a specific axial coordinate $z$ by integrating over the whole dataset $I(x_0,x)$ at a $z$-dependent angle $\theta(z)$; this allows one to perform, in post-processing, a software $z$-scanning similar to the hardware scan done by manually moving the focus of a conventional (incoherent imaging) device. 

The relation between the integration angle and the reconstructed axial plane can be understood by considering that the object point at coordinate $\bm x^\prime$, once illuminated by the source lit at transverse coordinate $\bm x_0$, is mapped onto the detector coordinate $\bm x$, which depends on both $\bm x_0$ and $\bm x^\prime$. As anticipated, the geometrical locus of the points of the sensor $\bm x$ corresponding to the same object coordinate $\bm x^\prime$ is a line in the $(x_0,x)$ space with equation
\begin{equation}
s_{x^\prime}: \alpha(\delta)\, x_0 + \beta(\delta)\, x+x^\prime=0,
\label{eq:line}
\end{equation}
where $\alpha$ and $\beta$ are two functions depending on both the defocusing distance $\delta$ and the particular illumination scheme; as in conventional LF imaging, they are obtained through ray tracing in a geometrical optics context.
The same holds, with the same coefficients $\alpha$ and $\beta$, for the other two coordinates $(y_0,y)$.
Therefore, for an object imaged at an axial displacement $\delta$ from the focused plane, the most accurate reconstructed image is $R_{z=\delta}$, as obtained by performing the integration in Eq. \eqref{eq:Radon} along lines with a tilting $\theta=\arctan (-\alpha/\beta)$.

As we shall discuss in the ``Results'' section, the 3D performance of the proposed coherent 3D imaging technique can be characterized in terms of the image fidelity $F_A[R_z]$. However, volumetric reconstruction shall require the image fidelity to be evaluated onto a three-parameter space: since the focus of the system is fixed at a given coordinate, both the relative position $\delta$ of the $s$-sized object and the reconstruction coordinate $z$ can be moved independently with respect to the plane at focus.

\section{Results}

\begin{figure}
    \centering
    \includegraphics[width=\textwidth]{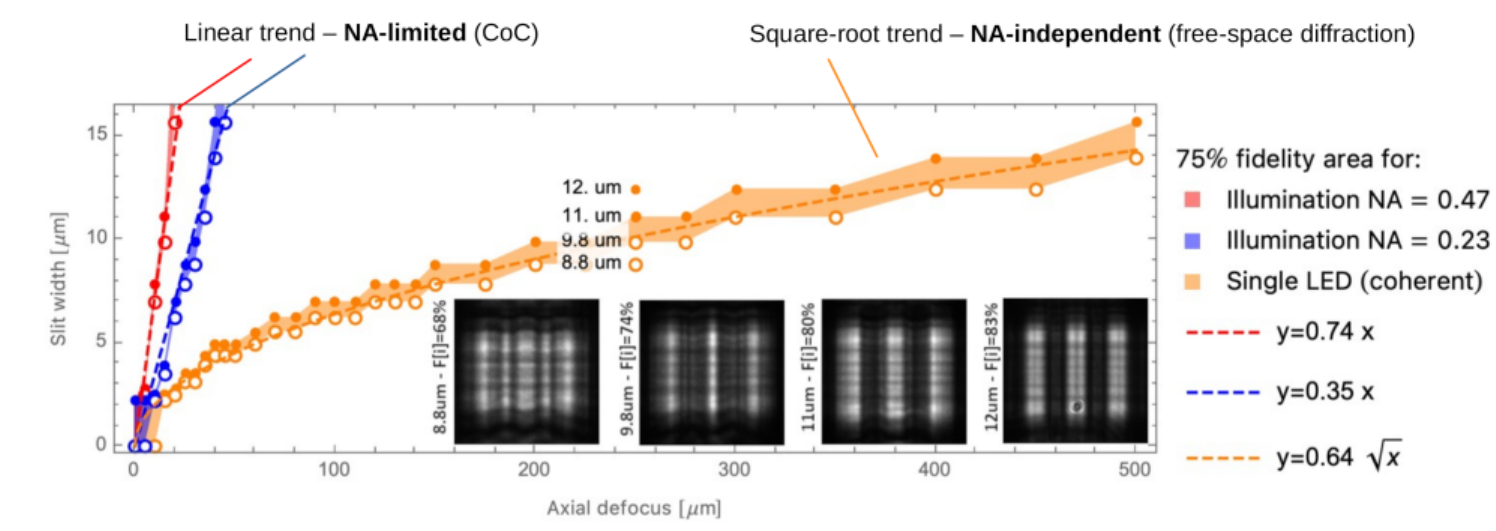}
    \caption{\label{fig:fig3}
    \textbf{Experimental comparison of resolution and DOF in incoherent and coherent imaging.}
    Comparison of the resolution versus DOF trade-off in two cases of incoherent imaging (red and blue points), corresponding to two different illumination NA, and in the case of coherent imaging (orange points), as obtained by single LED illumination. For both coherent and incoherent imaging, an image is considered to be resolved when the corresponding fidelity is at least $75\%$. The areas of $75\%$ fidelity (colored regions) are identified by the smallest slit width imaged with fidelity over $75\%$ (circles) and the largest slit width imaged with fidelity lower than $75\%$ (bullets). All data are taken with a conventional $20\times$ microscope, whose details are reported in the Supplementary document. The analysis is based on triple-slit masks from a USAF test target, having slit width $s$ and center-to-center slit distance $d=2s$. The four images, reported as a reference quality of the targeted fidelity, are taken at the same axial distance of $z=250\,\mu$m and correspond to the 4 different reported values of the slit width.
    }
\end{figure}

The introduction of the image fidelity has enabled to directly compare the performance of coherent and incoherent systems and to discover that, in coherent imaging, the degradation of the image resolution with defocusing is not related with geometrical blurring mechanisms such as the CoC. On the contrary, the degradation of the image quality is governed almost entirely by diffraction from the object plane to the imaged plane; this leads to a square-root law scaling $\sqrt{|z|}$ of the resolution with the distance from focus $z$, as opposed to the linear scaling characterizing the CoC.
Thus, in coherent imaging, the axial range in which the object is resolved thus scales quadratically with the object size, rather than linearly. Therefore, in addition to being independent of the NA of the imaging device, coherent imaging yields a quantitative DOF advantage over conventional (incoherent) imaging. 

In Fig.~\ref{fig:fig3}, we report the experimental demonstration of this prediction and show that coherent illumination enables a $4$ times larger DOF at $2\,\mu$m resolution and an almost $20$ times larger DOF at $10\,\mu$m resolution, with respect to incoherent illumination.
The experimental images shown in the figure are obtained by illuminating the masks with an array of green LEDs, placed $110\,$mm apart from the plane at focus of a $20\times$ magnification conventional microscope.
Although it is a well-known fact in microscopy, and in bright-field imaging in general, that source collimation indeed implies DOF augmentation \cite{microscopy_chap5}, we should highlight that the square-root trend we experimentally demonstrate is the result of an entirely different physical phenomenon and cannot be understood in terms of source collimation, but only in terms of its spatial coherence. 
The conventional (incoherent imaging) explanation of the DOF improvement through source collimation, in fact, is related to the divergence of the illuminating beam becoming smaller than the acceptance angle of the optical devices, so that the optical properties of the imaging device are no longer dictated by the NA of imaging device, but rather by the effective NA defined by the illumination itself. 
However, this effect is profoundly different from the DOF extension enabled by spatial coherence, where collimation is by no means a requirement.
The presented DOF advantage, in fact, is by all means maintained even with a quasi-infinite illumination NA, as one could get by bringing the illumination stage in extreme proximity to the sample and employing smaller sources, such as quantum dots and single-molecule LEDs.

\begin{figure}
	\centering
    	\includegraphics[width=.80\textwidth]{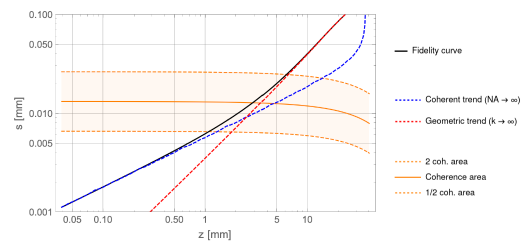}
	\caption{\textbf{Effect of partial coherence in the transition from coherent to incoherent imaging.}
    	\label{fig:partial}
    Log-log plot of the $95\%$-fidelity  (black continuous line) evaluated in the case of partial coherence due to the finite transverse size of the light source. 
    The dashed red and blue curves are the $95\%$-fidelities evaluated, respectively, in the geometric optics approximation and in wave optics but with infinte NA.
    All curves are obtained by considering a microscope with NA=$0.5$, illuminated by a green light emitter placed at $11\,$cm from its plane at focus and having a Gaussian intensity profile of width $w=0.3\,$mm; the object is a Gaussian slit of width $s$.
    The two dashed orange lines identify, for each defocusing $z$, the values of the slit width $s$ corresponding to $1/2$ (lower curve) to $2$ (upper curve) times the coherence area; the orange colored area identify the values of $s$ for which partial coherence enters into play giving rise to  the transition from incoherent to coherent imaging. 
    }
\end{figure}
The incoherent effects of DOF improvement, however, indeed exist in the regime in which the illumination collimation is such to define the effective NA of the system, but the coherence area on the sample is not wide enough for the system to behave in a coherent manner.
To gain more insight about the role played by the NA of the illumination system, we shall now study the transition from incoherent to coherent imaging and consider the general case of a finite-sized source, matching the experimental conditions of Fig. \ref{fig:fig3}.
In Fig.~\ref{fig:partial}, we plot the $95\%$-fidelity curve (solid black line) of the image of a transmissive mask (an $s$-sized slit) as a function of its distance from a $w$-sized \textit{incoherent} emitter; strictly speaking, $z$ is the distance of the object from the plane at focus, but its variation naturally changes the object-to-source distance as well. When the object is so close to the source that the spatial coherence acquired through propagation towards the sample is smaller than $s$, the resolution versus DOF trade-off is determined by the numerical aperture of the \textit{illumination}, as expected in a conventional system; this is demonstrated by the overlap of the evaluated fidelity (black line) with the one obtained in the geometrical optics approximation (red dashed line), for large values of $z$. In this regime, imaging is thus incoherent. 
However, as the object is moved farther away from the source, the coherence area on the object becomes proportionally larger till coherence effects become dominant, and the fidelity trend (black line) detaches from the geometrical optics prediction and overlaps on the coherent trend (dashed blue line), completely NA-independent. 
The yellow region highlights the transition from the incoherent to the coherent imaging, and shows that coherent effects enter into play, as predicted, when the coherence area becomes comparable to the details one wishes to resolve. 
Although this result is quite intuitive, its implications are very relevant. 
First, it demonstrates that the maximum DOF that direct imaging can achieve is ultimately limited, at least in the realms of classical optics, by the spatial coherence of the illumination on the sample; this is in contrast with the approximately infinite DOF one might incorrectly expect by interpreting the case of perfectly collimated illumination in terms of conventional incoherent imaging, along the line of what discussed above. 
Second, it leads to better appreciate the implications of the approach employed for obtaining the results of Fig.~\ref{fig:fig3}, namely, the reduction of the illumination stage area for showing, on the one hand, the effect of the NA on the CoC of incoherent imaging (blue and red points), on the other hand, the transition from incoherent to coherent imaging (orange points). The trend of the blue and red points indicate that the decrease of the illumination NA by a factor of $2$ increases the DOF by the same amount, as expected for incoherent imaging; however, in both cases, the illumination area (\textit{i.e.}, the radius of the area of lit LEDs) is not yet small enough for the coherence area to be comparable with the desired resolution. By further shrinking the illumination, a point is reached where the linear trend is lost and the of NA-independent square-root trend emerge, as a consequence of the larger coherence acquired by the illumination. The transition is explained in terms of spatial coherence.

\begin{figure}
    \centering
    \includegraphics[width=.9\textwidth]{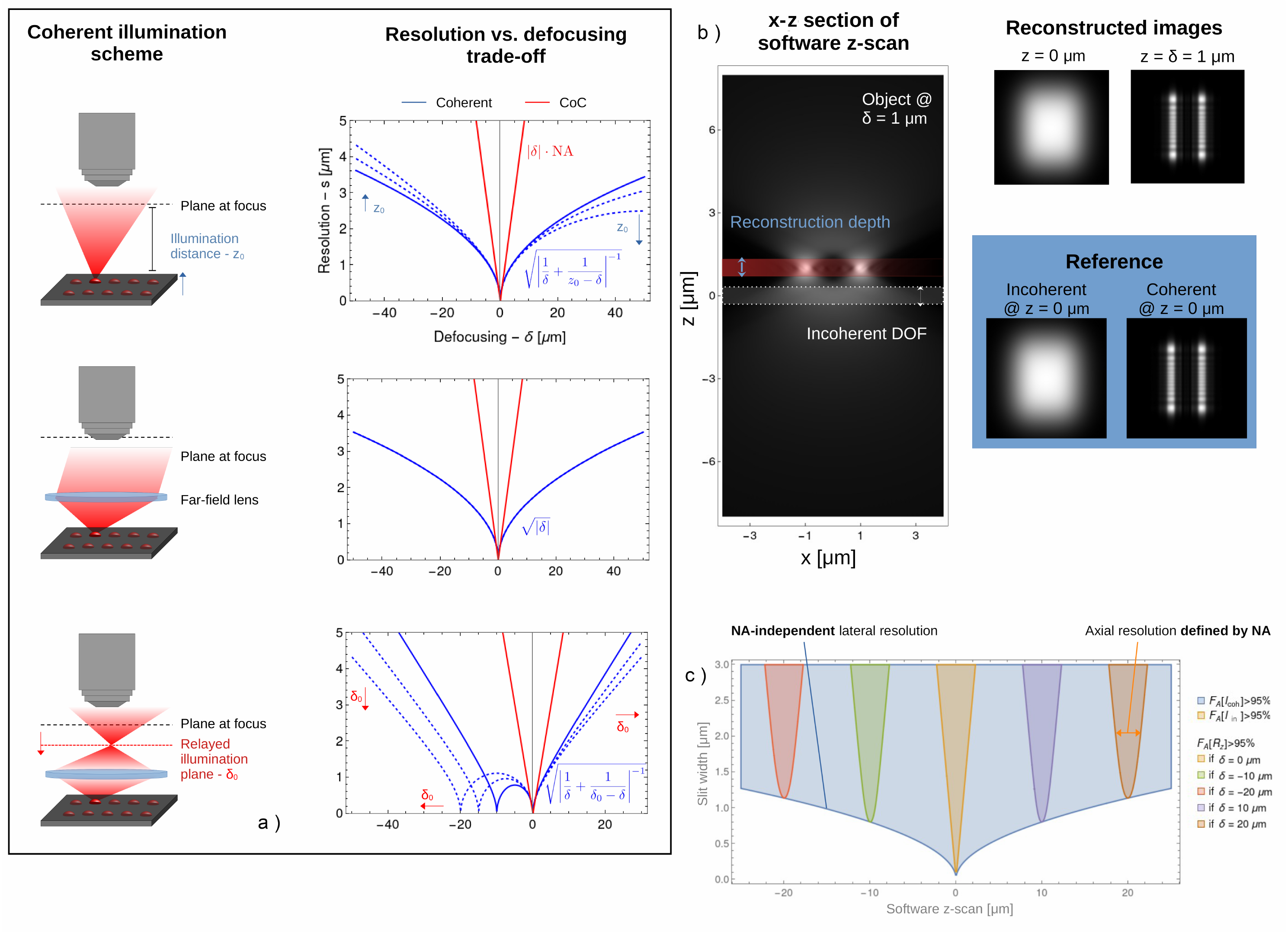}
    \caption{\label{fig:fig5}
    \textbf{3D imaging capability of a high-NA microscope exploiting spatial coherence.}
    \textit{Panel a).} Schematic representation of three possible illumination schemes for performing 3D imaging through coherent illumination. The plots on the right hand side report the corresponding scaling of the resolution as a function of the defocusing $\delta$, as prescribed by the coherent scaling $\sqrt{d(\lambda)}$ (blue curves); the different scaling obtained within a given scheme for varying values of the object-to-illumination distance ($z_0$ in the first case, $\delta_0$ in the third case) are also reported (dashed blue lines).
    \textit{Panel b).} Axial section of a stack of reconstructed images $R_z(\bm x)$ of a double slit mask placed at $\delta = 1\,\mu$m (left), showing the capability of localizing the sample. The two images reconstructed at the focused distance $z=0$ and at the correct sample locations $z=\delta=1\,\mu$m are reported in the top right panels, and compared with the corresponding two reference images reported in the bottom right panels: the defocused images collected by the same imaging device in the cases of incoherent (left) and coherent (right) illumination (same images reported in Fig.~\ref{fig:fig1}). The illumination scheme adopted for this simulations is the upper one of panel a), with an illumination distance $Z_0=110\,$mm.
    \textit{Panel c).} Characterization of the resolution as a function of the defocusing $z$ for five different object positions $\delta$, showing the sectioning capability and the overall resolution versus DOF performance of the proposed approach. The yellow and blue area are a reference for the performance of incoherent and coherent imaging, respectively; the last representing the maximum achievable DOF of the proposed technique. The five ``V-'' shaped areas show the sectioning capability enabled by the software z-scanning and the characterize the performance of the reconstructions, for the five different object placements; the software z-scan for a focused object (yellow) is shown to give the same resolution versus DOF performance as incoherent imaging.
    }
\end{figure}
\vspace{.1cm}Let us now show why localized coherent illumination is so convenient for performing 3D scanning-free imaging. As detailed in the ``Methods'' section, 3D information about the sample is gathered by measuring the 4D function $I(\bm x_0,\bm x)$, obtained by sequentially illuminating the object from different point-source on the illumination plane. The image dataset $I(\bm x_0,\bm x)$ can then be Radon-transformed to isolate axial planes of the samples, as prescribed by Eq. \eqref{eq:Radon}. 
In particular, an extremely interesting result is obtained by applying the Radon reconstruction to the plane $\delta$ on which an object with transmission function $\mathcal A$ is placed, namely:
\begin{equation}
    R_\delta(\bm x)=\left\lvert
        \mathcal A(\bm x)*\tilde{\mathcal P}(\bm x)
    \right\rvert^2,
    \label{eq:reconstruction_good}
\end{equation} 
with 
\begin{equation}
    \tilde{\mathcal P}(\bm x) = \mathcal D_{d(\delta)}(\bm x) * \mathcal P_0(\bm x),
\end{equation}
where $\mathcal P_0$ is the \textit{coherent} PSF of the imaging system in its focus, and $\mathcal D_z$ is propagation in vacuum by a distance $z$, as in Eq.~\eqref{eq:PSF}.
The properties of the reconstructed image are easily understood by noticing that Eq.~\eqref{eq:reconstruction_good} is exactly the expression of a coherent image (see Eqs.~\eqref{eq:coh_vs_inc} and \eqref{eq:PSF}), as observed by the same imaging device, but affected by an \textit{equivalent defocusing} $d(\delta)$. More specifically, coherent illumination gives rise to images having a resolution at focus defined by the Rayleigh limit and a resolution out-of-focus scaling with the square root of the defocusing $\sqrt{\delta}$, whereas the images reconstructed by our method have the same resolution at focus, but an out-of-focus resolution scaling with $\sqrt{d(\delta)}$.
As reported in Fig.~\ref{fig:fig5}a), different illumination schemes are thus possible, each one characterized by a specific scaling of the resolution with the defocusing; the optical performance of the device can thus be greatly enlarged, offering a wide flexibility in view of a variety of different applications. 
In fact, the plots indicate that the scaling of the lateral resolution as a function of the defocusing is a pure square-root only in the case of plane-wave illumination, namely, when $Z_0\rightarrow \infty$, in the first scheme, or when the middle scheme is adopted. 
In the other cases, the scaling remains defined by a square-root law, but the actual dependence also involves the illumination distance ($Z_0$ in the first scheme, $\delta_0$ in the third one).

Fig.~\ref{fig:fig5}b) demonstrates that the great (NA-independent) DOF extension typical of coherent imaging is integrated with very accurate (NA-dependent) axial localization, due to the incoherent imaging properties brought in by the reconstruction process. 
The reported software $z$-scan shows that a double-slit object placed outside of the native DOF of the microscope can actually be localized extremely well around the plane where the most accurate reconstruction happens. Also, at a glance, one immediately recognizes that the \textit{depth} of the reconstruction is not what one would expect by coherent imaging, but rather exactly the native (NA-defined) incoherent DOF of the device. As shown on the right hand side of the axial scanning, however, the reconstructed image is exactly the same image that coherent imaging would give, with an object displaced by an equivalent defocusing $d(\delta)$.

All the aforementioned properties are summarized by Fig.~\ref{fig:fig5} c). The blue and yellow areas identify the resolution performance of coherent and incoherent illumination, respectively, with their characteristic linear (CoC) and square-root trends. The colored ``V''-shaped regions, instead, show the optical properties of the images reconstructed, through software z-scanning, by the proposed 3D imaging modality, and correspond to five different axial positions of the object. As expected from our findings, the depth of the reconstructions, which represents the axial resolution of the 3D imaging technique, coincide with the DOF of incoherent imaging, namely, it is the same one would obtain by focusing the equivalent incoherent imaging system ({\it i.e.}, be lighting on the whole array of small sources at once) on the correct object plane; the only difference with respect to the image obtained by mechanical z-scanning within an equivalent incoherent imaging system is in the minimum resolution, which in our approach lies on the square-root curve defined by coherent imaging. 
On the other hand, when the object is placed at focus ($\delta=0$), the image reconstructed with our method is exactly the same as in incoherent imaging, both in the minimum Rayleigh-limited resolution and in the CoC-defined axial localization.

\begin{figure}
    \centering
    \includegraphics[width=\textwidth]{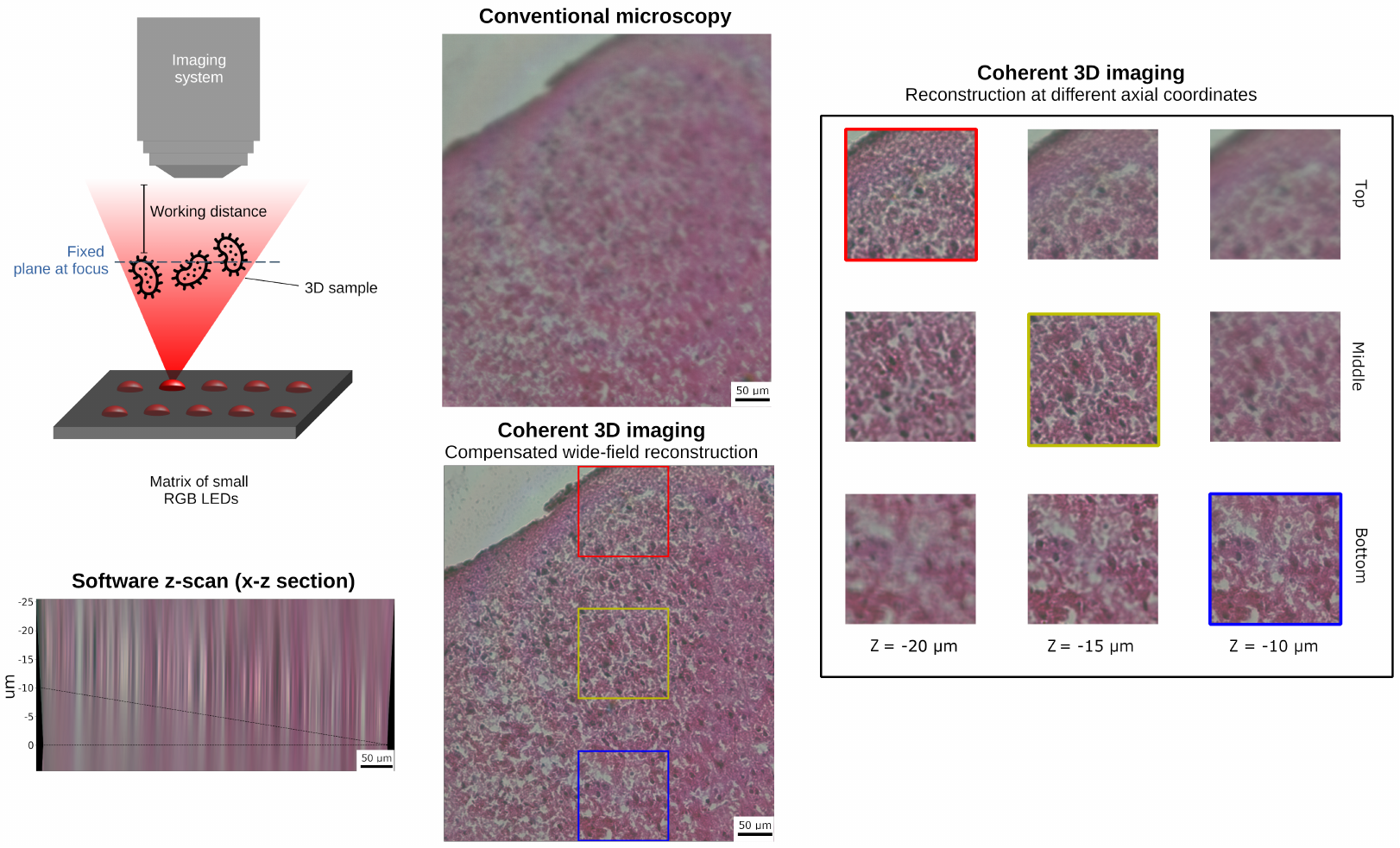}
    \caption{\label{fig:fig6}
    \textbf{Multicolor 3D reconstruction of a histological mouse brain section exploiting spatially coherent illumination} 
    The resolution versus DOF advantage granted by spatially coherent illumination is used to reconstruct the true-color wide-field image a histological mouse brain section marked at two wavelengths.
    Due to both a slight tilting of the sample holder and a axial misplacement, conventional incoherent illumination yields an unfocused image.
    By exploiting sequential coherent illumination from an array of RGB LEDs, volumetric information is obtained and employed, through the software axial scans, for reconstructing the different portions of the sample at different $z$ coordinates. The 3D information is then used to compensate for the tilting of about $10$ degrees and obtain a single wide-field image at focus.
    The optical microscope is a conventional $20\times$, $0.75$ NA wide-field device.
    }
\end{figure}

We shall now employ these results to experimentally demonstrate the high-resolution volumetric multicolor capability of the proposed technique (Fig.~\ref{fig:fig6}). Coherent illumination from localized emitters is obtained through an array of commercial RGB LEDs, placed far enough from the sample plane for the coherence area on the sample plane to be comparable with the details of interest.
The sample is a $10\,\mu$m-thick mouse brain section, where cell nuclei and cytoplasm have been labeled, respectively, by hematoxylin and eosin.
The acquired 3D information enables to clearly compensate for the sub-optimal placement of the microscope slide, whose closest part to focus is $10\,\mu$m away from the focused plane, and mounted with a tilting of about $10$ degrees.
Unlike the color-independent CoC, the square-root scaling of the resolution of coherent imaging has a weak dependence on wavelength ($\propto\sqrt\lambda$), thus giving rise to images characterized by negligible chromatic aberration.

\section{Discussion}

We have found that the lateral resolution and DOF of defocused images obtained through spatially coherent illumination are decoupled from the numerical aperture of the imaging system.
Such independence is particularly convenient for designing 3D imaging devices exploiting transverse coherence of light: the resulting overall DOF of the technique becomes independent of the optical components used for the image acquisition, and is instead entirely defined by the coherent illumination scheme, as shown in Fig.~\ref{fig:fig5}. 
Despite being based on the same image reconstruction principle as LF imaging, our resolution scaling is much more convenient, with the additional benefit of retaining Rayleigh-limited resolution at focus. In fact, compared to conventional LF devices, which achieve DOF extension at the expense of the lateral resolution, 3D imaging systems based on spatially coherent illumination have a DOF that scales quadratically with the desired resolution, thus always yielding an advantage over the linear scaling typical of LF~\cite{8924770}.
Furthermore, since a large NA has no effect on the resolution and DOF of the system, large apertures can be used to obtain optimal sectioning capability upon refocusing, enabling a strong suppression of the background neighbouring planes, as in high-NA tomographic systems.

Since our proposal only requires transverse spatial coherence, these systems work with \textit{temporally} incoherent sources, which induce negligible to modest radiation damage, as required by \textit{in vivo} biological applications.
Although image reconstruction through Radon transform does not recover the phase content, as opposed to computational techniques based on coherence~\cite{Tian:15, FourPtych_2021, alma9926533469005776}, it carries the enormous advantage of being performed in real-time with current GPU architectures and FPGAs \cite{Kertsz2016ANM}, or through the use of holographic screens~\cite{Wen:19}.
The proposed 3D wide-field imaging technique can thus be used both for direct and real-time imaging.

The extreme simplicity and low cost of the optical design, also compared to LF imaging, has high potentials to open up the possibility of employing 3D imaging in new scenarios, low-budget applications as well as for public healthcare in developing countries.

\begin{backmatter}
\bmsection{Funding}
All Authors acknowledge funding from Università degli Studi di Bari through the Horizon Europe Seeds program, project INTERGLIO (S081). M.D. and F.V.P. are supported by PNRR MUR project PE0000023-NQSTI. 
M.D., G.M. and F.V.P. are supported by INFN project QUISS.
MD, GM and FVP acknowledge funding under project ADEQUADE: this project has received funding from the European Defence Fund (EDF) under grant agreement EDF-2021-DIS-RDIS-ADEQUADE (n°101103417).
G.P.N. is supported by: 1) AstroDyn (FA9550-19-1-0370), AstroColl (FA9550-21-1-00352) and Stochastic Biophysical Interactions within Aquaporin-4 Assemblies (FA9550-20-1-0324) funded by AFOSR; 2) Marie Skłodowska-Curie Actions -ITN-2020 ASTROTECH (GA956325) funded by the European Commission; 3) NEXTGENERATIONEU (NGEU) funded by the Ministry of University and Research (MUR), National Recovery and Resilience Plan (NRRP), project MNESYS (PE0000006) – A Multiscale integrated approach to the study of the nervous system in health and disease (DD 1553, 11.10.2022); 4) NEXTGENERATIONEU (NGEU) funded by the Ministry of University and Research (MUR), National Recovery and Resilience Plan (NRRP), project CN00000041 - National Center for Gene Therapy and Drugs based on RNA Technology (DD n.1035, 17.06.2022).\\
\textbf{Note.} Funded by the European Union. Views and opinions expressed are however those of the author(s) only and do not necessarily reflect those of the European Union or the European Commission. Neither the European Union nor the granting authority can be held responsible for them.

\bmsection{Author contributions}
G.M. conceptualized the idea, developed theory and simulations, designed and built the experimental setup, developed the software, performed data analysis, and wrote the original draft. G.M. and B.B. performed the experiments. B.B. prepared the histological sample and contributed to write the biological part in the Supplemental document. F.V.P. supervised the theoretical work. G.P.N. supervised the biological part of the work. M.D. supervised the overall physical part of the work. M.D. and G.S. contributed to the organization and writing of the manuscript
M.D. and G.P.N. were responsible for fundings. All authors read and edited the manuscript.

\bmsection{Disclosures}
The authors declare no competing interests.

\bmsection{Data Availability}All the data leading to the results discussed in the paper are available upon
request to the corresponding author.

\bmsection{Supplemental document}
See Supplementary text for supporting content. 

\end{backmatter}

\bibliography{bibliography}






\clearpage

\end{document}